%% file: main.tex
\documentclass{article}

\input{sections/_preamble}

\input{sections/_header}
\begin{document}
\maketitle
\input{sections/abstract}

\input{sections/introduction}

\input{sections/relwork}

\input{sections/methods}
\input{sections/results}

\input{sections/discussion}

\bibliographystyle{plain} % We choose the "plain" reference style
\bibliography{refs}

\input{sections/appendix/main}

\end{document}

%% file: sections/_preamble.tex
\usepackage{longtable}
\usepackage{graphicx} % Required for inserting images
\usepackage{booktabs}
\usepackage{hyperref}
\usepackage{here}
\usepackage{xcolor}
\usepackage{natbib}
\usepackage{cleveref}
\usepackage{pdflscape}
\usepackage[a4paper, total={6in, 10in}]{geometry}

\usepackage{amssymb}% http://ctan.org/pkg/amssymb
\usepackage{pifont}% 
\usepackage{caption}
\usepackage{subcaption}
\usepackage{CJKutf8}
\usepackage{comment}

\newcommand{\zh}[1]{\begin{CJK*}{UTF8}{gbsn}#1\end{CJK*}}

\newcommand{\cmark}{\checkmark} % ✓
\newcommand{\xmark}{\textcolor{gray}{$\times$}}
\newcommand{\Npapers}{$65{,}241$}
\newcommand{\Nlit}{$41$}

\newcommand{\NsampleGT}{$80$}

\newcommand{\nGSCC}{$8{,}105$}

\newcommand{\NnamesChinese}{$81{,}177$}
\newcommand{\NnamesPinyin}{$68{,}601$}

\newcommand{\NauthorsChinese}{$106{,}411$}
\newcommand{\NauthorsPinyin}{$99{,}222$}

\newcommand{\SimThrEng}{$0.9621$}
\newcommand{\SimThrCh}{$0.9207$}

\newcommand{\NnamesChineseTwoPapers}{$35{,}873$}
\newcommand{\NnamesPinyinTwoPapers}{$33{,}441$}

\newcommand{\precisionCh}{$1.0$}
\newcommand{\precisionEng}{$0.953$}
\newcommand{\recallEng}{$0.82$}
\newcommand{\recallCh}{$0.80$}
\newcommand{\recallAvg}{$0.81$}
\newcommand{\fscoreCh}{{$0.89$}}
\newcommand{\fscoreEng}{{$0.88$}}

\newcommand{\crossLangPerfectOverlap}{{$87.3\%$}}
\newcommand{\crossLangAvgJaccard}{$0.9220$}

%% file: sections/_header.tex
\title{Bridging the Language Gap in Scholarly Data \textsc{i}: Enhancing Author Disambiguation Algorithms for Chinese Names}
\author{Mingrong She$^{1}$\footnote{\texttt{m.she@maastrichtuniversity.nl}}, Liuhuaying Yang$^{2}$, Ana Maria Jaramillo$^{2,3}$, Lisette Espín-Noboa$^{2,3}$
\\
\footnotesize $^1$Maastricht University, Maastricht, The Netherlands\\
\footnotesize $^2$Complexity Science Hub, Vienna, Austria\\
\footnotesize $^3$Graz University of Technology, Graz, Austria 
}

\date{}

%% file: sections/abstract.tex
\begin{abstract}
Disambiguating scholars with identical names is essential for accurate authorship assignment and robust large-scale scientometric research.
Existing methods are often designed for Latin-script metadata and perform poorly on Chinese names.
In international publications, Chinese names typically appear as Romanized Pinyin (e.g., ``Wang Wei''), which is highly ambiguous as it can map to multiple distinct characters (e.g., \zh{王伟}, \zh{王威}, \zh{王维}).
Chinese characters, in contrast, reduce but do not eliminate this ambiguity, and are rarely available in international records.
To address both challenges, we propose a rule-based disambiguation framework that integrates co-authorship networks, citation networks, author affiliations, and content similarity.
We apply this framework to \Npapers~physics papers from the China National Knowledge Infrastructure (CNKI), spanning over $70$ years of data. %1953--2024.
On a human annotated sample of \NsampleGT~name pairs, our method achieves F1-scores of \fscoreEng~for Pinyin names and \fscoreCh~for character-based names, outperforming two baseline approaches, with improvements driven primarily by higher recall.
The comparable performance across both writing systems shows that our approach is script-agnostic, enabling reliable large-scale scientometric analyses.
\end{abstract}

%% file: sections/introduction.tex
\begin{figure}[t]
    \centering
    \includegraphics[width=1.0\linewidth]{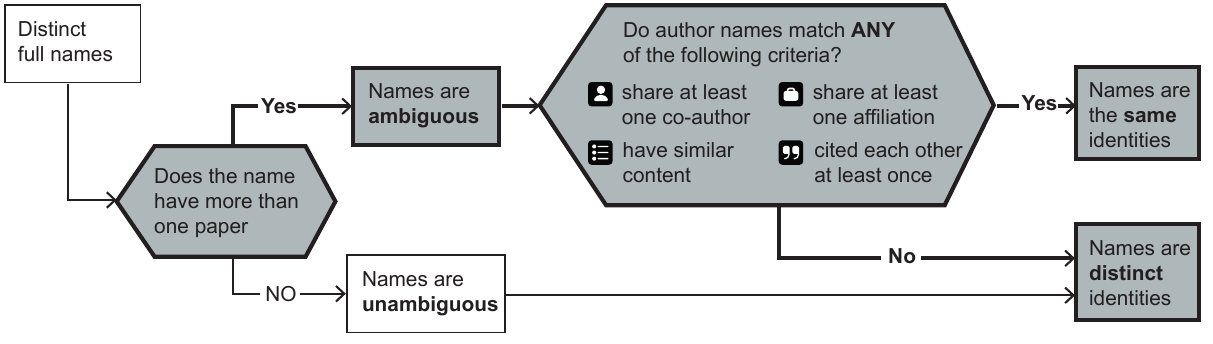}
    \caption{
    \textbf{Overview of the proposed rule-based framework for resolving author name ambiguity.}
    A distinct name associated with multiple papers is considered ambiguous.
    To resolve this ambiguity, our framework evaluates co-authorship, affiliation, citation links, and content similarity: if any criterion matches, the author names are merged as the same identity; otherwise, they are assigned to distinct identities. The disambiguation process is highlighted in dark gray.}
    \label{fig:methodology}
\end{figure}

\section{Introduction}
Author name disambiguation is a critical task in scientometric research, playing a crucial role in analyzing scholarly collaboration, citation networks, and academic career trajectories. 
Accurate attribution of academic works to their rightful authors is essential to ensure the reliability of research evaluations and to assess academic productivity~\cite{Manzoor2022Toward, Schulz2016Using}. 
However, name ambiguity poses a significant challenge, particularly in large academic databases where multiple researchers may share identical or similar names~\cite{aksnes2008different}. 
Misattribution of authorship can lead to errors in measuring academic influence and distort evaluations of the impact of a researcher over time~\cite{Schulz2016Using, kim2016distortive, harzing2015health}.
This challenge is especially pronounced for Chinese names, where the rapid growth of Chinese authors~\cite{zhang2018china, yuan2018international} has expanded the pool of researchers sharing identical names, heightening ambiguity and demanding better disambiguation methods~\cite{kim2023effect,xu2024rethinking}.

Several structural properties of Chinese names exacerbate this problem.
First, Chinese family names are highly concentrated. 
Approximately $2,000$ Han Chinese family names are currently in use, but the top $100$ account for around 87\% of the population, dramatically increasing the likelihood of author-name collisions~\cite{louie2008chinese, granshaw2019research, qiu2008identity,xu2024rethinking}.
Second, Chinese full names typically lack middle names or initials, eliminating a common source of differentiation in the Western context~\cite{kim2023effect, sun2017name}. 
Third, Chinese given names draw from a vast set of characters with no predefined pool of common names. 
In international databases, these names are transliterated into Pinyin, a process that introduces significant ambiguity. 
Multiple characters can map to the same Pinyin syllable, and a single character may correspond to different pronunciations~\cite{halpern2016some, teixeira2020chinese}. 
For example, the Pinyin name ``Wang Wei'' can correspond to \zh{王伟}, \zh{王威}, or \zh{王维}, among others. As a result, authors with distinct Chinese names often appear identical in Pinyin.
Finally, Chinese full names place the family name first, whereas English full names place it last, complicating the parsing of name strings in databases that assume Western conventions.
These properties together limit the effectiveness of traditional disambiguation methods, which are primarily optimized for Western names.

In this paper, we propose a rule-based disambiguation algorithm for Chinese names, designed to address the language-specific and transliteration-related challenges described above.
We build upon the algorithm of Sinatra et al.~\cite{sinatra2016quantifying} and extend it with the semantic similarity component proposed by Waqas and Qadir~\cite{waqas2021multilayer}.
For each distinct name, our method collects all associated papers to determine whether they represent the same identity (\Cref{fig:methodology}). 
Distinct names with only one paper are resolved as a single identity.
The remaining distinct names are ambiguous. For each, we compare all pair of papers and evaluate the author names against four conditions:
(1) they share a co-author;
(2) they share an institutional affiliation;
(3) one cites the other's work; or
(4) their publications are semantically similar.
Two author names satisfying any condition are resolved as the same identity; otherwise, they are distinct. 
The first three conditions capture direct social and institutional ties; content similarity complements them by detecting thematic proximity when such ties are absent.

We apply our algorithm to \Npapers~academic papers from Chinese physics journals indexed by the China National Knowledge Infrastructure (CNKI)\footnote{\url{https://cnki.net}}, from which we extract author names, co-authors, affiliations, citations, and abstracts.
To assess the impact of transliteration, we run the algorithm independently on two versions of the dataset: one containing the original Chinese text, and another with names converted to Pinyin and titles and abstracts translated into English.
Across the full corpus, we identify \NnamesChinese~(\NnamesPinyin) distinct names and resolve them into \NauthorsChinese~(\NauthorsPinyin) identities using Chinese characters (Pinyin).
We evaluate accuracy on a human annotated sample of 80 name pairs drawn from this corpus.
All three methods, ours and the two baselines of Sinatra et al.~\cite{sinatra2016quantifying} and Waqas \& Qadir~\cite{waqas2021multilayer}, achieve comparably high precision, rarely misclassifying two distinct identities as the same individual.
Our method outperforms both baselines in recall, identifying same-identity pairs that the baselines fail to detect. %, largely due to the addition of abstract similarity.
This gain in recall drives higher overall F1-scores: 0.89 for Chinese character-based names and 0.88 for Pinyin names.
The comparable performance across both writing systems shows that our approach is script-agnostic, with 87\% of identity assignments agreeing between the two representations across the full corpus.

In summary, this paper makes the following contributions:
\begin{enumerate}
    \item We propose a rule-based disambiguation algorithm that extends existing social and institutional relation-based methods with content similarity, and demonstrate that this addition improves recall without sacrificing precision.
    \item We apply our algorithm to both native Chinese records (character names, Chinese content) and international-database representations (Pinyin names, English-translated content), demonstrating that disambiguation accuracy remains high in both cases despite the increased name ambiguity introduced by romanization.
    \item We make available upon request a dataset of \Npapers~CNKI physics papers with dual-language metadata and disambiguated author identities.
    \item To facilitate transparency and reproducibility, we make all code and analysis scripts publicly available at~\cite{she2026chinesenames}.
\end{enumerate}

%% file: sections/relwork.tex
\section{Related work}
\label{sec:related_work}
Name disambiguation has attracted sustained attention in bibliometrics, digital libraries, and information retrieval~\citep{smalheiser2009author,ferreira2012brief}. To map Chinese-specific approaches, we conducted a systematic search of the Clarivate Web of Science database (see~\Cref{app:sec:literature_search} for the full protocol). This search yielded 41 relevant papers spanning 1998--2023, summarized in~\Cref{app:tbl:name_disambiguation_literature_search}. We organize our review around methodological approaches and three critical cross-cutting issues that expose key gaps in the literature and motivate our approach.

Existing Chinese name disambiguation methods can be broadly categorized into four approaches. \emph{Supervised and semi-supervised methods} such as probabilistic models~\cite{WOS:000208170400002,WOS:000302946800002}, graph convolutional networks~\cite{WOS:000800267300006}, and multi-kernel functions with external verification~\cite{WOS:000385137600006} demonstrate strong performance but require substantial labeled data, exhibit topic sensitivity, and face scalability challenges. \emph{Unsupervised clustering approaches}~\cite{WOS:000297823300051,WOS:000425306300001,WOS:000674763700007} 
automatically group authors based on similarity metrics computed from metadata such as co-authors but do not leverage semantic or contextual information from publication content. \emph{Heuristic and rule-based methods}~\cite{WOS:000307730000009,WOS:000344638800007} apply predefined matching rules (e.g., name and institution matching) and are computationally efficient, but are limited by their reliance on a small set of features. More recently, \emph{semantic and network-based approaches} such as semantic fingerprinting~\cite{WOS:000401747900036} and dual-channel heterogeneous graph networks~\cite{WOS:000700631800001} have emerged to better capture contextual author relationships, but depend heavily on metadata richness and typically require substantial labeled corpora that are not available for Chinese-domestic databases. 

Our setting instead calls for a method that requires no labeled data and can be applied without retraining to both Chinese-character and Pinyin representations of the same corpus---criteria that point toward heuristic, feature-based approaches. The closest precedents come from outside the Chinese disambiguation literature. Sinatra et al.~\cite{sinatra2016quantifying} propose a rule-based algorithm that identifies candidate author pairs by matching last names together with first names (identical or sharing an initial), then resolves ambiguity by checking whether the pair shares at least one co-author, shares at least one institutional affiliation, or one cites the other's work. This approach is data-efficient and has been widely applied to physics bibliographies, but it does not exploit the semantic content of publications: two records with no social or institutional tie cannot be resolved even when their research is demonstrably similar. Waqas \& Qadir~\cite{waqas2021multilayer} address this limitation by supplementing social and institutional ties with additional similarity signals: email addresses, publication venues, titles, abstracts, and keywords. Neither method is designed for Chinese names or evaluated on Pinyin-transliterated data, leaving open how well such heuristic approaches transfer to Chinese-specific disambiguation settings.

These observations point to three key gaps that motivate our study. 
(i)~Most Chinese name disambiguation methods rely solely on direct social and institutional ties, without exploiting publication semantics. The two most relevant heuristic approaches---Sinatra et al. and Waqas \& Qadir---were developed on Western bibliographic data and have not been evaluated on Chinese metadata, let alone on parallel character and Pinyin versions of the same corpus. 
(ii)~Insufficient attention has been paid to how name representation affects disambiguation accuracy. Most studies focus on a single script~\cite{WOS:000351310600002,WOS:000385137600006,WOS:000401747900036,WOS:000401380300007,WOS:000425306300001,WOS:000511927600001,WOS:000674763700007,WOS:000800267300006}, and only one~\cite{WOS:000657876300001} has conducted a direct comparison between Chinese characters and Pinyin. 
This gap is particularly significant because international databases predominantly index names in Pinyin while domestic Chinese databases use Chinese characters.
(iii)~There is a critical lack of transparency and reproducibility: many studies do not describe the databases used for evaluation, and open-source implementations remain extremely rare, with only 2 of 41 papers (4.9\%) releasing accessible code. 

To address these gaps, we propose a rule-based disambiguation algorithm that combines the social and institutional tie criteria of Sinatra et al.~\cite{sinatra2016quantifying}---co-authorship, affiliations, and citations---with the content similarity component introduced by Waqas \& Qadir~\cite{waqas2021multilayer}. Like Sinatra et al., our method requires no labeled training data and is directly applicable to new domains and databases; the addition of abstract similarity extends its coverage to author pairs linked thematically but not socially or institutionally (gap~i). We apply the full pipeline to a large Chinese-domestic (CNKI) physics corpus in both Chinese-character and Pinyin forms, providing the first systematic script-agnostic evaluation on a large-scale Chinese dataset (gap~ii). All code, data, and analysis scripts are publicly released to support reproducibility (gap~iii).

%% file: sections/methods.tex
\section{Data and methods}
Our methodology comprises three main components designed to systematically evaluate the impact of language representation on Chinese author name disambiguation. First, we construct parallel datasets containing identical academic papers in both original Chinese and translated English forms, enabling direct comparison of disambiguation performance across language representations. Second, we implement a multi-dimensional disambiguation algorithm that integrates co-authorship networks, institutional affiliations, content similarity, and citation relationships. Finally, we conduct a comparative evaluation using human-annotated ground truth to assess algorithm performance on both Chinese and English datasets. This design allows us to quantify how transliteration affects disambiguation accuracy while controlling for all other variables.

\subsection{Chinese character dataset}
We compiled a comprehensive dataset from the China National Knowledge Infrastructure (CNKI), the world's largest repository of Chinese academic literature. %information, 
Our analysis focused on $20$ journals published by the Chinese Physical Society over ten years from 1953 to 2024 (further details in~\Cref{app:sec:data}). Physics was selected due to its rich publication history, large author communities, and comparability with international datasets such as APS publications.\footnote{\url{https://journals.aps.org/datasets}} For each article, we extracted metadata, including titles, author lists, authors' affiliations, publication years, abstracts, keywords, and the journals in which they were published. This systematic collection yielded \Npapers~papers encompassing \NnamesChinese~distinct Chinese character full names, providing a robust foundation for the disambiguation analysis. % were initially retrieved for analysis.
It is important to note that this dataset contains only author names as originally assigned to papers, without disambiguation labels. To enable evaluation, we constructed a smaller ground-truth dataset via expert annotation (see~\Cref{sec:gt}).

\subsection{English translation dataset}
Chinese author names in international bibliometric databases are typically represented in Pinyin rather than Chinese characters. To assess whether our disambiguation algorithm remains robust across different linguistic representations of the same content, we constructed an English translation dataset. This dataset construction involved two main steps: converting Chinese author names to Pinyin and translating textual content to English.

For the first step, we constructed a Pinyin list for all Chinese characters. 
We avoided using Google Translate because it is designed for general sentence-level translation and often performs poorly with proper nouns such as Chinese personal names. Python-based Pinyin tools face similar issues: some Chinese characters have name-specific pronunciations that differ from their standard readings. And these tools fail to resolve polyphonic characters accurately, leading to mistranslation or inconsistent transliteration across contexts.
Instead, we used \nGSCC~General Standard Chinese Characters from HanziDB\footnote{\url{http://hanzidb.org/character-list/general-standard}} and identified their Pinyin pronunciations via the online dictionary Hanzi Quanxi,\footnote{\url{https://qxk.bnu.edu.cn/\#/}} which provides standardized name-specific pronunciations and allows systematic handling of polyphonic characters. 
We excluded characters that only have neutral tone pronunciations (e.g., \zh{吗, 呢, 啊}), as these characters are rarely used in personal names.
For characters with multiple pronunciations, we prioritized the pronunciation typically used in given names or family names. 

For the second step, we used Google Translate to translate titles, abstracts, and keywords from Chinese to English, producing an English dataset with the same key textual elements. To verify that the translation preserves semantic content, we computed pairwise similarity scores within each corpus separately: for every pair of papers, we obtained one similarity score from the Chinese corpus and one from the English corpus. We used cosine similarity between Word2Vec document embeddings (see \Cref{app:sec:similarity_calculation}), which measures the angle between document vectors and is therefore insensitive to document length---making it appropriate for comparing papers of varying lengths. Comparing the two sets of scores, one per corpus, we found a Pearson correlation of $0.89$ ($p < 0.001$), indicating that paper pairs that are semantically similar in Chinese tend to be similarly related in English. This supports the reliability of the translated dataset for disambiguation purposes.

\subsection{Ground-truth dataset}
\label{sec:gt}
To evaluate the performance of our disambiguation algorithm, we constructed a ground-truth dataset with human verified labels. We randomly selected 80 pairs of Chinese character names: 40 pairs predicted by the algorithm to belong to the same individual and 40 pairs predicted to belong to different individuals, despite having the same names. The same ground-truth labels were applied to the Pinyin pairs for comparative evaluation between the two linguistic representations.

For each pair of names, annotators were provided with two publications, each including an author’s name and a title. They were asked to determine whether the two names referred to the same individual (detailed annotation interface and examples are provided in~\Cref{app:sec:annotation}).
To ensure reliability, two native Chinese speakers independently reviewed all 80 pairs. They agreed on 70 cases, disagreed on 4, and one annotator could not verify 6 names. The inter-annotator agreement in this first round was Cohen's kappa $= 0.75$ and Krippendorff's alpha $= 0.76$ (moderate agreement). The 10 cases in which initial agreement was not reached were re-examined and resolved through discussion between the annotators to reach consensus.

\subsection{Proposed disambiguation algorithm}
\label{sec:algo}

We employ a comprehensive disambiguation algorithm that leverages metadata from authors and publications. 
Our approach extends two existing methods, which we later use as benchmarks: 
Sinatra et al.~\cite{sinatra2016quantifying}, who use co-authors, affiliations, and citations, and 
Waqas \& Qadir~\cite{waqas2021multilayer}, who incorporate publication metadata such as title, abstract and keywords. 
We integrate both approaches and apply them to the same dataset in two linguistic representations, mirroring how Chinese authors appear in domestic versus international databases: 
native Chinese (character names, Chinese content) and romanized English (Pinyin names, English-translated content).

The algorithm operates on distinct names associated with at least two publications (\Cref{fig:methodology}). 
For each pair of authors with identical names, it evaluates four matching criteria. % in sequence. 
A pair is classified as the same identity as soon as any criterion is satisfied. Distinct names that remain unmatched after all four stages are classified as different identities.
The first criterion is \textbf{shared co-authorship}. Two author names are considered the same identity if they share at least one co-author across their publication records. Researchers tend to collaborate repeatedly with the same colleagues, making co-authorship a strong disambiguation signal.
For author names not resolved by co-authorship, the algorithm next evaluates \textbf{institutional affiliation similarity}. Two author names are matched if their affiliation strings achieve a normalized Levenshtein similarity ratio~\cite{yujian2007normalized} above $0.6$. They also match if one affiliation string contains the other after removing non-alphanumeric characters.
The third criterion computes the cosine \textbf{similarity between papers} using Word2Vec-based~\cite{mikolov2013efficient} document vectors constructed from titles, abstracts, and keywords. A pair is matched when this similarity exceeds a language-specific threshold: \SimThrCh~for Chinese and \SimThrEng~for English. Separate thresholds are required because the Chinese-character and English corpora occupy different embedding spaces---Word2Vec vectors built from Chinese text yield systematically different similarity score distributions than those built from English text, thus a single cutoff cannot be applied uniformly across both representations. Each threshold was determined through a grid search to maximize consistency between the two linguistic representations by minimizing both proportional overlap differences and classification discrepancies; the detailed selection procedure is described in~\Cref{app:sec:similarity_calculation}. These conservative thresholds reduce the risk of false positives.
The fourth and final criterion examines \textbf{citation relationships}. Two author names are considered the same identity if one has cited papers by the other, suggesting self-citation.

\subsection{Baseline implementation}
To evaluate our extensions, we implement the two foundational methods described above. 
For Sinatra et al.~\cite{sinatra2016quantifying}, we apply the original algorithm using affiliation, citation, and coauthor networks. 
For Waqas \& Qadir~\cite{waqas2021multilayer}, we adapt their approach to our available metadata, using coauthor networks, affiliations, publication titles, abstracts, and keywords, but excluding author name variants, email addresses, and publication venues, which are not present in our CNKI dataset.

%% file: sections/results.tex
\section{Results}

\begin{figure}[t]
\includegraphics[width=1\textwidth]{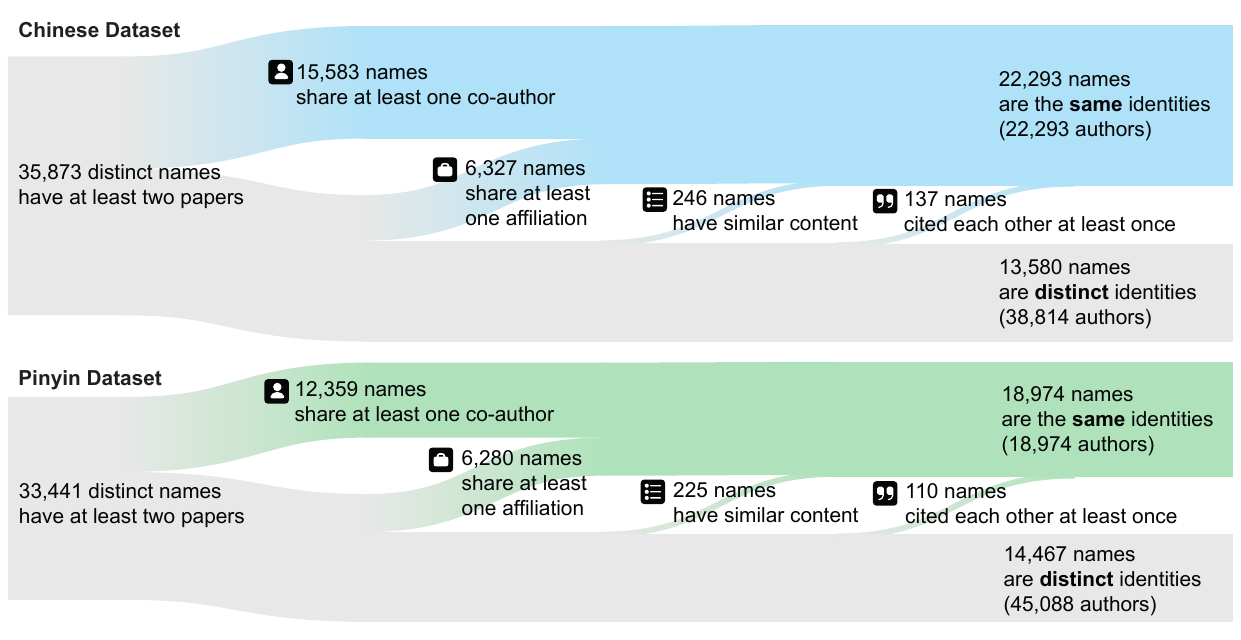}
\caption{\textbf{Name disambiguation data-flow in Chinese and English datasets}. 
The diagram presents the sequential filtering outcomes of the author name disambiguation algorithm applied separately to the Chinese dataset, where author names appear as Chinese characters (top panel), and the English translation dataset, where author names appear in Pinyin romanization (bottom panel). 
Each panel begins with all \textit{distinct names} associated with at least two publications and proceeds through four matching criteria: shared co-authorship, affiliation similarity, content similarity, and citation links. The numbers within each box indicate the count of \textit{distinct names} satisfying each criterion, ultimately resolved into disambiguated author \textit{identities}.}
\label{fig:pipeline_results}
\end{figure}

\subsection{Disambiguation outcomes}

We applied the disambiguation algorithm separately to the Chinese character dataset and the English translation dataset (containing Pinyin-transliterated author names and English-translated paper content).
The Chinese dataset contains \NnamesChinese~distinct names, of which \NnamesChineseTwoPapers~(44\%) appear on  at least two publications. 
The English dataset contains \NnamesPinyin~distinct Pinyin names, of which \NnamesPinyinTwoPapers~(49\%) appear on at least two publications (\Cref{fig:pipeline_results}). Only these ambiguous distinct names---those with multiple publications---enter the disambiguation pipeline.
Shared co-authorship, the first criterion, resolves the largest share of ambiguous names: 15,583 in the Chinese dataset and 12,359 in the English dataset. This result indicates that co-authorship networks carry a strong signal for disambiguating authors.
Affiliation similarity identifies 6,327 additional matches in the Chinese dataset and 6,280 among the English dataset. Content similarity captures 246 and 225 further matches. Citation relationships, the final criterion, resolve 137 and 110 remaining cases.

After all four stages, 22,293 distinct Chinese character names and 18,974 distinct Pinyin names are each attributed to a single identity.  The remaining 13,580 and 14,467 distinct names correspond to multiple identities---38,814 and 45,088, respectively  (\Cref{fig:pipeline_results}). Combined with the single-publication author names, the pipeline resolves \NauthorsChinese~total identities in the Chinese character dataset and \NauthorsPinyin~in the English dataset.

\begin{figure}[t]
\centering
\includegraphics[width=0.8\textwidth]{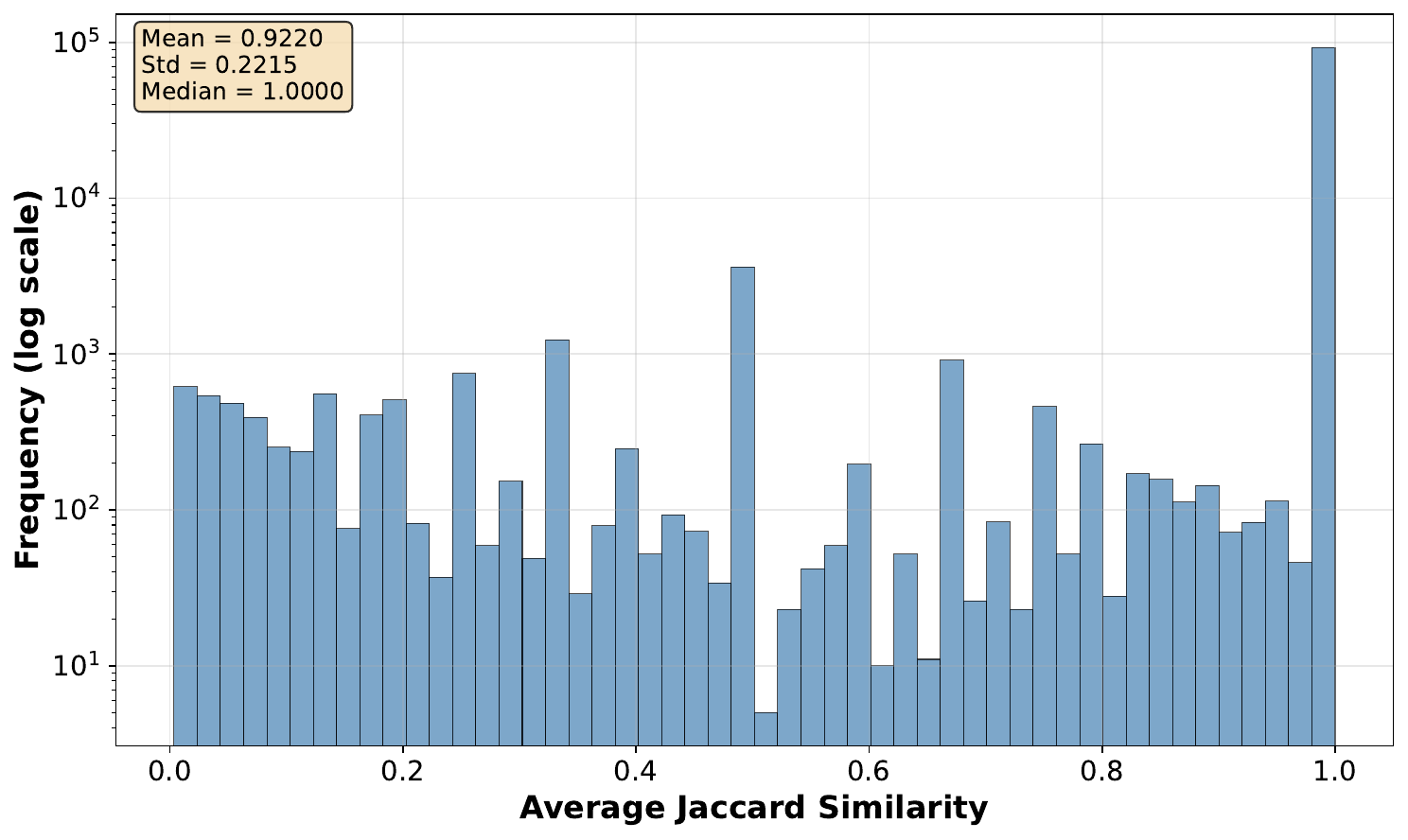}
\caption{\textbf{Cross-language disambiguation consistency.} 
Distribution of average Jaccard similarity between \NauthorsChinese~Chinese identities and matching Pinyin identities (measuring overlap in assigned papers). 
The concentration at 1.0 (87.3\%) indicates identical disambiguation across languages.
}
\label{fig:jaccard_distribution}
\end{figure}

\subsection{Chinese vs. English disambiguation agreement}

Because the Chinese character and English datasets represent the same underlying  publications, the two disambiguation results should produce consistent identity  groupings. 
We measure this consistency using the Jaccard similarity between  each identity's paper set across the two representations.

For each identity in the Chinese character dataset, we converted its name to Pinyin and identified all matching identities with the same Pinyin name in the English dataset. We computed the Jaccard similarity between the paper sets of each Chinese identity and its matching Pinyin identities, then averaged the scores across all matches.
Across the \NauthorsChinese~Chinese identities, the mean Jaccard similarity is \crossLangAvgJaccard~(\Cref{fig:jaccard_distribution}). 
The distribution is skewed towards perfect agreement. 
A total of \crossLangPerfectOverlap~of identities achieve a score of $1.0$, indicating that the algorithm assigns them each identical paper sets under both representations. 
The remaining 12.7\% show a wide spread of scores, with some near zero (no overlap). 
These discrepancies likely reflect the many-to-one mapping from Chinese characters to Pinyin, which creates additional name collisions in the English dataset and splits identities differently. 
Overall, the algorithm is largely stable across representations, though Pinyin-induced ambiguity affects a meaningful share of identities.

\subsection{Evaluation}

\begin{figure}[t]
\includegraphics[width=1\textwidth]{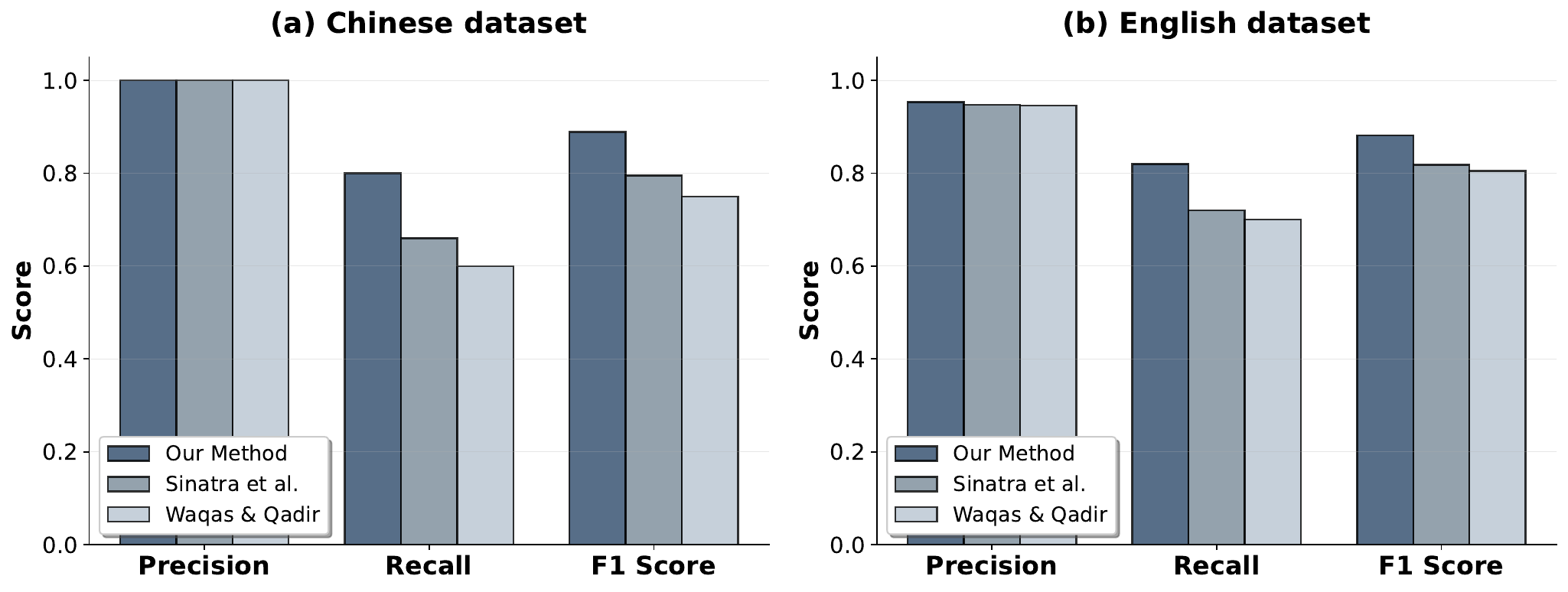}
\caption{\textbf{Disambiguation performance comparison.} 
Our method outperforms existing baselines (Sinatra et al. and Waqas \& Qadir) across all metrics on both (a) Chinese and (b) English datasets, with particularly strong gains in recall.}
\label{fig:Comparison_Sinatra_Waqas}
\end{figure}

We evaluate our algorithm against the human-annotated ground-truth dataset (see \Cref{sec:gt}) using precision, recall, and F1-score, benchmarking it against two existing approaches: Sinatra et al.~\cite{sinatra2016quantifying} (SI) and Waqas \& Qadir~\cite{waqas2021multilayer} (WQ). \Cref{fig:Comparison_Sinatra_Waqas} summarizes the results.

Our method outperforms both baselines in both linguistic representations with F1-scores of \fscoreCh~(Chinese) and \fscoreEng~(English), demonstrating stable performance across them. 
SI shows similar stability with slightly lower scores (0.80 Chinese, 0.82 English), while WQ exhibits a larger improvement from Chinese to English (0.75 to 0.80).
Examining precision and recall separately reveals the sources of these performance differences. 
All three methods achieve perfect precision (1.0) on the Chinese dataset, indicating no false merges of distinct authors. 
However, precision drops slightly in the English dataset to \precisionEng~for our method, with SI and WQ showing comparable degradation. Our method's advantage lies primarily in recall. 
We achieve \recallCh~(Chinese) and \recallEng~(English), maintaining stable performance across representations while consistently outperforming both baselines.
SI shows lower recall (0.66 Chinese, 0.72 English), while WQ achieves the lowest recall overall (0.60 Chinese, 0.70 English), both showing moderate cross-representation variation.
Across all methods, recall is higher on the English dataset than on the Chinese dataset.
This indicates that all approaches fail to merge some author names that should be unified in the Chinese representation, though this problem is more pronounced for the baselines.

Our recall advantage stems from incorporating content similarity as a fourth matching criterion, which captures author pairs that co-authorship, affiliation, and citation signals alone (used by SI) fail to identify. This gain in recall comes without a meaningful loss in precision, confirming that the conservative similarity thresholds (\Cref{app:sec:similarity_calculation}) effectively prevent false positives. 
The consistent improvement across both representations further supports the robustness of our approach. 
Detailed confusion matrices are provided in \Cref{app:sec:Confusion_Matrices}.

%% file: sections/discussion.tex
\section{Discussion}
Author name ambiguity in Chinese bibliographic data manifests through two distinct error types. 
Understanding these patterns reveals both the limitations of current disambiguation methods and pathways for improvement.

\subsection{Consequences of ambiguous names}
Author name disambiguation errors take two forms, each affecting bibliometric reliability differently. 
The first involves incorrectly merging distinct individuals into a single author profile. 
Our Chinese character dataset achieves perfect precision (\precisionCh, \Cref{fig:Comparison_Sinatra_Waqas}a), demonstrating that native-script representations fully avoid this error. 
The English dataset using Pinyin names achieves high but imperfect precision at \precisionEng~(\Cref{fig:Comparison_Sinatra_Waqas}b), reflecting how romanization loses the distinctiveness preserved in Chinese characters. 
Conflation errors inflate citation counts and distort collaboration networks~\cite{kim2016distortive, harzing2015health, strotmann2012author}. 
For example, in 2011, ``Y. Wang'' was identified as the most prolific author in scientific literature, credited with 11 papers per day across multiple disciplines, a physically impossible rate from merging hundreds of distinct individuals~\cite{xu2024rethinking}. 
Such errors undermine institutional rankings and funding decisions that rely on author-level indicators.

The second error type involves fragmenting a single scholar's work across multiple author profiles. 
Both our Chinese and English datasets achieve recall of approximately \recallAvg, substantially outperforming baseline methods but indicating room for improvement. 
Fragmentation occurs when authors lack distinctive publication signatures, particularly for early-career researchers with sparse co-authorship networks and limited topical consistency, or established scholars who have shifted research domains and whose newer work shows low textual similarity to earlier publications~\cite{Schulz2016Using}.
Fragmentation systematically disadvantages individual researchers by scattering their citation counts and obscuring their publication trajectories~\cite{strotmann2012author, kim2016distortive}. 
At a systemic level, both error types distort bibliometric analyses in different directions: 
conflation artificially concentrates credit while fragmentation disperses it, creating a dual problem where some author profiles are over-counted and others under-counted. 
This is particularly consequential for scholars from linguistic backgrounds where name ambiguity is most severe, creating representational inequities in global bibliometric systems used for research evaluation and cross-national collaboration studies~\cite{aksnes2008different, xu2024rethinking}.

\subsection{Limitations and future directions}
Our dataset comprises papers from Chinese physics journals and may not generalize to disciplines with different structural characteristics. 
Humanities scholars, for instance, more frequently publish single-authored work with distinct citation dynamics compared to engineers~\cite{praus2025note}. 
Testing our approach across fields with varying collaboration norms would reveal whether discipline-specific adjustments are necessary. 
Our validation confirms our method's effectiveness but was conducted on a limited sample. 
Larger-scale human evaluation would provide stronger evidence of generalizability across author populations and time periods.

While our method improves upon existing baselines, unresolved fragmentation points to the limits of text-based similarity measures alone. 
Two complementary solutions merit consideration. 
First, integration of ORCID identifiers would provide definitive author attribution~\cite{xu2024rethinking}. 
However, ORCID adoption remains incomplete, particularly among senior researchers who established their careers before persistent identifiers became standard, inactive scholars no longer maintaining digital profiles, and researchers in disciplines with lower adoption rates~\cite{porter2025understanding}. 
Until universal adoption is achieved, algorithmic disambiguation remains necessary. 
Second, encouraging Chinese scholars to provide their names in both Chinese characters and standardized Pinyin transliterations when submitting manuscripts would reduce ambiguity at the source. 
This dual-representation approach would preserve the high precision we observe in Chinese character datasets while maintaining accessibility for international databases. 
Publishers and repositories could facilitate this by creating structured metadata fields for multiple name representations~\cite{smith2020transitioning}. 

Similar disambiguation challenges affect other non-Western naming systems (Korean, Japanese, Vietnamese, Indian)~\cite{sungwon2018disambiguation,kurakawa2014researcher}, suggesting our approach could generalize beyond Chinese with appropriate language-specific adaptations.
Expanding our validation to include Web of Science and Scopus would enable cross-database performance assessment and reveal whether indexing practices create systematic biases. 
Future methodological refinements could incorporate additional metadata when available, citation profile analysis that examines not only whom authors cite but also the topical composition of cited papers, and temporal weighting schemes that relax similarity thresholds for papers separated by a few years to account for topic drifts~\cite{zeng2019increasing}.
These enhancements would address the fragmentation problem while maintaining the high precision our current approach achieves.

\section{Conclusion}

We present a rule-based disambiguation framework that achieves robust performance for Chinese author names across native and romanized English representations. 
Applied to \Npapers~physics papers from CNKI, the method achieves F1-scores of \fscoreCh~and \fscoreEng~respectively, outperforming established baselines in recall while maintaining high precision. 
The framework demonstrates strong cross-language stability, with \crossLangPerfectOverlap~of identities showing perfect agreement across both representations. 
This consistency matters because Chinese researchers publishing in international databases risk having their identities fragmented across systems or conflated under shared Pinyin transliterations, distorting collaboration networks and productivity metrics. 
The openly released code~\cite{she2026chinesenames} and dataset (available upon request) enable direct application to both Chinese-domestic and international corpora, providing infrastructure for more accurate tracking of scholarly contributions. 
Future work should extend this framework to other East Asian writing systems and explore integration with persistent identifier systems to bridge domestic and international scholarly records.

%% file: sections/appendix/main.tex
\clearpage 
\cleardoublepage
\appendix
%%%%%%%%%% Insert Supplementary Material here %%%%%%%%%%%%%%
\setcounter{figure}{0}
\renewcommand{\thefigure}{S\arabic{figure}}
\setcounter{section}{0}
\renewcommand{\thesection}{S\arabic{section}}
\setcounter{table}{0}
\renewcommand{\thetable}{S\arabic{table}}
\renewcommand{\theHtable}{S\arabic{table}}
\renewcommand{\theHfigure}{S\arabic{figure}}
\renewcommand{\theHsection}{S\arabic{section}}

\section*{Supplementary Information}
\input{sections/appendix/literature}
\input{sections/appendix/data}
\input{sections/appendix/semanticsim}

\input{sections/appendix/ground_truth}
\input{sections/appendix/confusion}

%% file: sections/appendix/literature.tex
\section{Systematic literature search}
\label{app:sec:literature_search}
To identify prior work on Chinese name disambiguation, we conducted a systematic search in the Clarivate Web of Science (WOS) database on 2023-10-22.
We used the following query, applied to the title, abstract, and full text of all indexed articles:

\begin{quote}
\texttt{ALL=(Chinese name disambiguation) OR ALL=(Chinese Personal Name Disambiguation) OR ALL=(Chinese author Name Disambiguation)}
\end{quote}

\noindent
This search returned \Nlit~articles published between 1998 and 2023, indicating sustained interest in the topic over more than two decades.
We manually screened all \Nlit~articles and excluded 21 that did not address name disambiguation (e.g., papers that mentioned the search terms only in passing or focused on unrelated problems).
The remaining 20 papers form the basis of our literature review in~\Cref{sec:related_work}.

We classified each relevant paper along six dimensions:
(i)~\textit{method}, categorized as supervised/semi-supervised, unsupervised clustering, heuristic/rule-based, or semantic/network-based (categories are not mutually exclusive, as some papers combine multiple approaches);
(ii)~\textit{dataset availability}, whether the authors shared their experimental data;
(iii)~\textit{code availability}, whether the authors released implementation code;
(iv)~\textit{Chinese character support}, whether the method was designed for or evaluated on names in Chinese characters;
and (v)~\textit{Pinyin support}, whether the method addressed Pinyin-transliterated names.
\Cref{app:tbl:name_disambiguation_literature_search} presents the results of this classification.

Among the 20 relevant papers, supervised or semi-supervised methods are most common (nine papers), followed by unsupervised clustering (six), heuristic or rule-based matching (five), and semantic or network-based approaches (three).
Nine papers focused on Chinese character names, four addressed Pinyin names, and only one~\citep{WOS:000657876300001} compared both representations.
Reproducibility remains a concern: only three papers shared datasets~\citep{WOS:000401380300007,WOS:000545966200002,WOS:000700631800001}, and only two released code~\citep{WOS:000344638800007,WOS:000700631800001}.

\begin{table}[H]
    \centering
    \caption{\textbf{Chinese name disambiguation literature.} 
Overview of \Nlit~articles retrieved from the Clarivate Web of Science database on 2023-10-22. 
The Methods columns categorize papers by approach: Sup/Semi = Supervised/Semi-supervised methods; Unsup Clust = Unsupervised Clustering approaches; Heur/Rule = Heuristic/Rule-based matching; Sem/Net = Semantic/Network-based approaches.
Other columns: Data Avail = Dataset Available; Code Avail = Code Available; CN Supp = Chinese character Support; PY Supp = Pinyin Support.
A \cmark indicates that the article uses the method or addresses the feature listed in the corresponding column, \xmark indicates its absence, and "-" indicates papers unrelated to name disambiguation.}
\label{app:tbl:name_disambiguation_literature_search}
    \resizebox{\linewidth}{!}{
    \begin{tabular}{@{}r l c c c c c c c c c@{}}
    \toprule
    \# & Reference & \multicolumn{4}{c}{Methods} & \begin{tabular}[c]{@{}c@{}}Data\\Avail\end{tabular} & \begin{tabular}[c]{@{}c@{}}Code\\Avail\end{tabular} & \begin{tabular}[c]{@{}c@{}}CN\\Supp\end{tabular} & \begin{tabular}[c]{@{}c@{}}PY\\Supp\end{tabular} \\
    \cmidrule(lr){3-6}
    & & {\begin{tabular}[c]{@{}c@{}}Sup/\\Semi\end{tabular}} & {\begin{tabular}[c]{@{}c@{}}Unsup\\Clust\end{tabular}} & {\begin{tabular}[c]{@{}c@{}}Heur/\\Rule\end{tabular}} & {\begin{tabular}[c]{@{}c@{}}Sem/\\Net\end{tabular}} & & & & \\ \midrule
    1 & Li and Yip (1998)~\cite{WOS:000077762600004} & \xmark & \xmark & \cmark & \xmark & \xmark & \xmark & \xmark & \xmark \\
    2 & Lin et al. (2010)~\cite{WOS:000276057900022} & \xmark & \xmark & \cmark & \xmark & \xmark & \xmark & \xmark & \xmark \\
    3 & Tang et al. (2010)~\cite{WOS:000208170400002} & \cmark & \xmark & \xmark & \xmark & \xmark & \xmark & \xmark & \xmark \\
    4 & Strotmann and Zhao (2012)~\cite{WOS:000307730000009} & \xmark & \cmark & \cmark & \xmark & \xmark & \xmark & \xmark & \xmark \\
    5 & Chen et al. (2012)~\cite{WOS:000297823300051} & \xmark & \cmark & \xmark & \xmark & \xmark & \xmark & \xmark & \xmark \\
    6 & Tang et al. (2012)~\cite{WOS:000302946800002} & \cmark & \xmark & \xmark & \xmark & \xmark & \xmark & \xmark & \xmark \\
    7 & Chin et al. (2014)~\cite{WOS:000344638800007} & \xmark & \xmark & \cmark & \xmark & \xmark & \cmark & \xmark & \cmark \\
    8 & Jiang et al. (2015)~\cite{WOS:000351310600002} & \cmark & \cmark & \xmark & \xmark & \xmark & \xmark & \cmark & \xmark \\
    9 & Xu et al. (2016)~\cite{WOS:000385137600006} & \cmark & \xmark & \xmark & \xmark & \xmark & \xmark & \cmark & \xmark \\
    10 & Han et al. (2017)~\cite{WOS:000401747900036} & \xmark & \xmark & \xmark & \cmark & \xmark & \xmark & \cmark & \xmark \\
    11 & Youtie et al. (2017)~\cite{WOS:000412527000021} & \xmark & \xmark & \cmark & \xmark & \xmark & \xmark & \xmark & \xmark \\
    12 & Wang et al. (2017)~\cite{WOS:000401380300007} & \cmark & \xmark & \xmark & \xmark & \cmark & \xmark & \cmark & \xmark \\
    13 & Zhu et al. (2018)~\cite{WOS:000425306300001} & \xmark & \cmark & \xmark & \xmark & \xmark & \xmark & \cmark & \xmark \\
    14 & Yin et al. (2020)~\cite{WOS:000511927600001} & \cmark & \cmark & \xmark & \xmark & \xmark & \xmark & \cmark & \xmark \\
    15 & Xu et al. (2020)~\cite{WOS:000545966200002} & \cmark & \xmark & \xmark & \xmark & \cmark & \xmark & \xmark & \xmark \\
    16 & Ma et al. (2020)~\cite{WOS:000537222600002} & \xmark & \xmark & \xmark & \cmark & \xmark & \xmark & \xmark & \cmark \\
    17 & Fan and Li (2021)~\cite{WOS:000674763700007} & \xmark & \cmark & \xmark & \xmark & \xmark & \xmark & \cmark & \xmark \\
    18 & Sheng et al. (2021)~\cite{WOS:000800267300006} & \cmark & \xmark & \xmark & \xmark & \xmark & \xmark & \cmark & \xmark \\
    19 & Zheng et al. (2021)~\cite{WOS:000700631800001} & \xmark & \xmark & \xmark & \cmark & \cmark & \cmark & \xmark & \cmark \\
    20 & Kim et al. (2023)~\cite{WOS:000657876300001} & \cmark & \xmark & \xmark & \xmark & \xmark & \xmark & \cmark & \cmark \\
    21 & Yuh et al. (2006)~\cite{WOS:000238233300015} & - & - & - & - & - & - & - & - \\
    22 & Li et al. (2013)~\cite{WOS:000323377600010} & - & - & - & - & - & - & - & - \\
    23 & Harzing (2015)~\cite{WOS:000365130100050} & - & - & - & - & - & - & - & - \\
    24 & Mao and Lu (2017)~\cite{WOS:000399186200001} & - & - & - & - & - & - & - & - \\
    25 & Huang et al. (2018)~\cite{WOS:000424188400026} & - & - & - & - & - & - & - & - \\
    26 & Chai et al. (2018)~\cite{WOS:000451790000010} & - & - & - & - & - & - & - & - \\
    27 & Zeng et al. (2018)~\cite{WOS:000434944600001} & - & - & - & - & - & - & - & - \\
    28 & Huang et al. (2020)~\cite{WOS:000517111200016} & - & - & - & - & - & - & - & - \\
    29 & Zhao et al. (2020)~\cite{WOS:000564148300014} & - & - & - & - & - & - & - & - \\
    30 & Jiang et al. (2020)~\cite{WOS:000591803100001} & - & - & - & - & - & - & - & - \\
    31 & Liu et al. (2021)~\cite{WOS:000604483600001} & - & - & - & - & - & - & - & - \\
    32 & Barrena et al. (2021)~\cite{WOS:000697191700004} & - & - & - & - & - & - & - & - \\
    33 & Xu et al. (2021)~\cite{WOS:000601162500035} & - & - & - & - & - & - & - & - \\
    34 & Zhang (2021)~\cite{WOS:000629724700001} & - & - & - & - & - & - & - & - \\
    35 & Zhang et al. (2021)~\cite{WOS:000679278600016} & - & - & - & - & - & - & - & - \\
    36 & Wang et al. (2022)~\cite{WOS:000766560600001} & - & - & - & - & - & - & - & - \\
    37 & Song et al. (2022)~\cite{WOS:000848235600004} & - & - & - & - & - & - & - & - \\
    38 & Zhou et al. (2022)~\cite{WOS:000798360300001} & - & - & - & - & - & - & - & - \\
    39 & Duan et al. (2022)~\cite{WOS:000887081700001} & - & - & - & - & - & - & - & - \\
    40 & Zhang et al. (2022)~\cite{WOS:000845722100001} & - & - & - & - & - & - & - & - \\
    41 & Wang et al. (2023)~\cite{WOS:000952961300026} & - & - & - & - & - & - & - & - \\ \bottomrule
    \end{tabular}}
\end{table}

%% file: sections/appendix/data.tex
\section{CNKI dataset}
\label{app:sec:data}
We collected papers from the China National Knowledge Infrastructure academic platform, focusing on 20 physics journals published by the Chinese Physical Society up to 2024-02-14 (see~\Cref{app:fig:CPS_data}).

\begin{table}[H] 
\centering

\caption{\textbf{Summary of journals published by the Chinese Physical Society.} The table shows the subject focus of each journal, language (CH=Chinese, EN=English), and the total number of papers published in each journal up to 2024-02-14.}
\resizebox{\linewidth}{!}{
\begin{tabular}{l l c c r}
\toprule 
Journal Name & Subject & \multicolumn{2}{c}{Language} & Papers \\
\cmidrule(lr){3-4}
& & CH & EN & \\
\midrule 
Acta Physica Sinica &  General Physics, Condensed Matter Physics, Theoretical Physics
 & \checkmark &  & 19,730 \\
Chinese Journal of Atomic and Molecular Physics & Atomic and Molecular Physics
 & \checkmark & \checkmark & 5,743 \\
Chinese Journal of Chemical Physics & Chemical Physics, Molecular Dynamics, Quantum Chemistry
 &  & \checkmark & 2,559 \\
Chinese Journal of High Pressure Physics & High Pressure Physics, Condensed Matter Physics, Materials Science
 & \checkmark &  & 852 \\
Chinese Journal of Light Scattering & Optics, Light Scattering, Spectroscopy
  & \checkmark &  & 595 \\
Chinese Journal of Liquid Crystals and Displays & Liquid Crystals, Display Technology, Materials Physics
 & \checkmark &  & 482\\
Chinese Journal of Luminescence & Luminescence, Optoelectronics, Photophysics
 & \checkmark &  & 2719 \\
Chinese Journal of Magnetic Resonance & Magnetic Resonance, NMR, Medical Physics
 & \checkmark & \checkmark & 7\\
Chinese Physics B & Condensed Matter Physics, Quantum Physics, Materials Physics
 &  & \checkmark & 8,813 \\
Chinese Physics C & Nuclear Physics, Particle Physics, High Energy Physics
 &  & \checkmark & 2,515 \\
Chinese Physics Letters & General Physics, Rapid Communications
 &  & \checkmark & 1,513 \\
College Physics & Physics Education
 & \checkmark &  & 6,081 \\
Communications in Theoretical Physics & Theoretical Physics, Mathematical Physics, High Energy Theory
 &  & \checkmark & 1,039 \\
Journal of Chinese Electron Microscopy Society & Electron Microscopy, Materials Characterization
 & \checkmark &  & 499 \\
Journal of Chinese Mass Spectrometry Society & Mass Spectrometry, Analytical Physics, Atomic and Molecular Physics
 & \checkmark &  & 2,059 \\
Journal of Quantum Optics & Quantum Optics, Nonlinear Optics
 & \checkmark &  & 1,402 \\
Nuclear Physics Review & Nuclear Physics, Nuclear Structure, Nuclear Reactions
 & \checkmark &  & 1,741\\
Physics Teaching & Physics Education
 & \checkmark &  & 1,001 \\
Progress in Physics & Physics Review, Frontier Topics, Interdisciplinary Physics
 & \checkmark &  & 514 \\
Wuli (Physics)
 & General Physics, Popular Science, Science Communication
 & \checkmark &  & 5,379 \\
\bottomrule 
\end{tabular}}
\label{app:fig:CPS_data}
\end{table}

%% file: sections/appendix/semanticsim.tex
\section{Semantic similarity computation}
\label{app:sec:similarity_calculation}

Our disambiguation pipeline uses semantic similarity between scholarly documents as one of the conditions for deciding whether two author name records refer to the same individual: two records are linked only if, among other criteria, the papers associated with them are sufficiently similar in content. This section describes how we compute that similarity and how we calibrate the threshold that determines when two documents are considered content-equivalent.

To represent each paper, we concatenated its available textual components---title, keywords, and abstract---into a single unified text. When a paper was missing keywords or an abstract, we used only the available components, ensuring full coverage of the corpus regardless of data completeness.

We embedded these representations using word embedding techniques based on the Word2Vec model~\cite{mikolov2013efficient}. Because our corpus contains documents in two languages, we trained two separate continuous bag-of-words Word2Vec models with 200-dimensional vector spaces: one on the Chinese corpus and one on the English corpus of translated abstracts. Both models were configured with identical hyperparameters to produce comparable representations across languages.

To compute the similarity between two documents, each text was first tokenized using language-appropriate methods (Chinese word segmentation for Chinese texts, standard tokenization for English texts). Each token was then mapped to its vector in the corresponding language model, with out-of-vocabulary tokens discarded. We computed the arithmetic mean of the resulting token vectors to obtain a single document-level embedding, then normalized each document vector to unit length to ensure comparisons are scale-invariant. Semantic similarity between a pair of documents was finally measured as the cosine similarity between their document vectors, yielding a value between 0 and 1.

With similarity scores in hand for all document pairs, the remaining task is to determine the threshold that will decide whether a pair of papers is considered semantically similar. Setting it presents two key challenges. First, the same pair of papers does not necessarily receive the same cosine similarity score across corpora---a pair might score 0.90 in Chinese but only 0.80 in English, so a single shared threshold would treat the two corpora inequitably. Second, and more fundamentally, we did not need the thresholds to be numerically equal; we needed them to agree on which pairs of papers are semantically similar. That is, the set of pairs classified as similar under the Chinese threshold should correspond as closely as possible to the set classified as similar under the English threshold. We therefore sought a pair of corpus-specific thresholds---one for the Chinese character corpus and one for the Pinyin transliteration corpus---that jointly maximize this cross-corpus consistency.

\begin{figure}[t]
\includegraphics[width=1\textwidth]{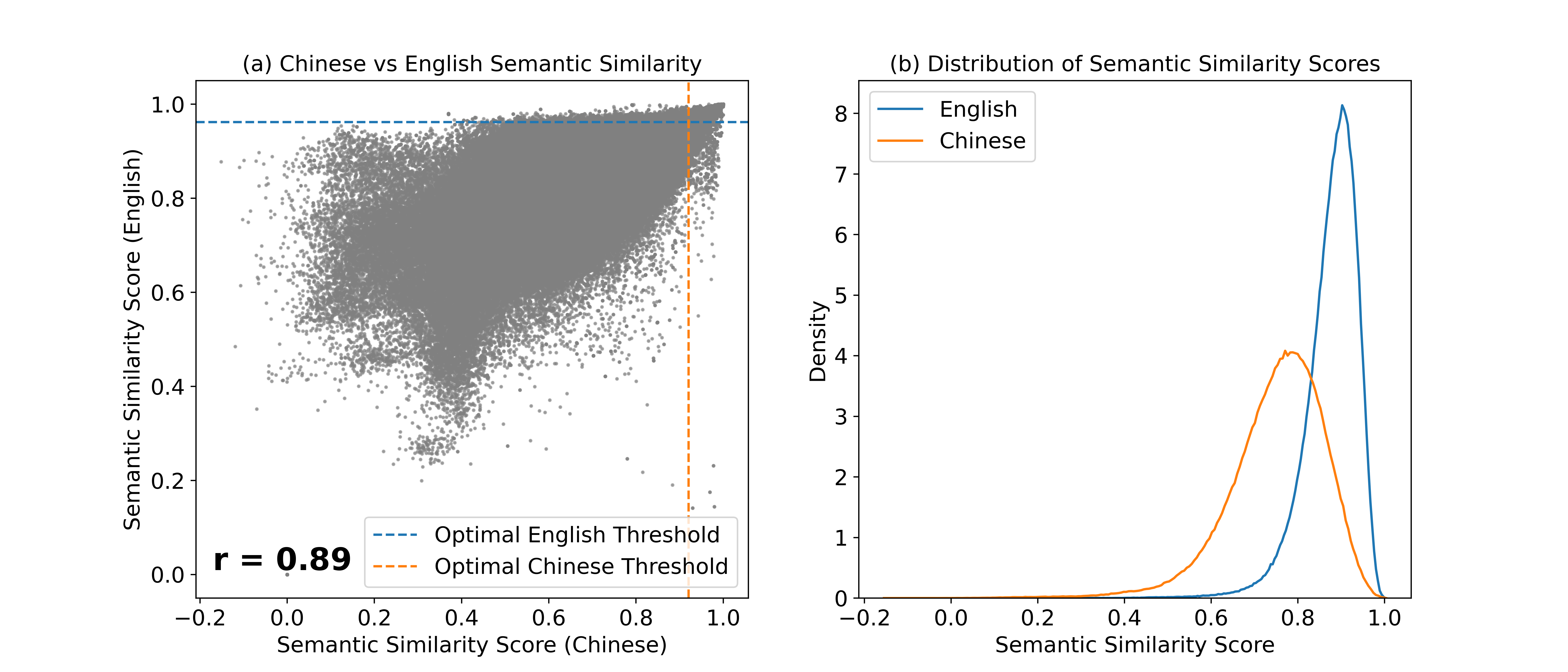}
\caption{\textbf{Semantic similarity scores across Chinese and English corpora.} Each point in panel (a) represents a pair of papers, plotted by their cosine similarity score in the Chinese corpus (x-axis) against their score in the English corpus (y-axis). The two measures are highly correlated (Pearson coefficient $r$ = 0.89). The vertical orange dashed line marks the optimal Chinese threshold (\SimThrCh) and the horizontal blue dashed line marks the optimal English threshold (\SimThrEng); pairs in the upper-right quadrant are classified as semantically similar in both corpora. Panel (b) shows the density distributions of similarity scores for each corpus (non-zero values only), illustrating that English scores are concentrated near 1.0 while Chinese scores are more broadly distributed, motivating the use of separate thresholds.}
\label{fig:SimilarityScatter}
\end{figure}

To find these thresholds jointly, we computed pairwise cosine similarity for all document pairs within each corpus and then conducted a grid search over the similarity spectrum from 0.9 to 1.0, evaluating 900 threshold combinations. For each candidate pair of thresholds we measured two quantities: (1) the difference in the proportion of pairs classified as similar in each corpus, and (2) the number of pairs that received inconsistent classifications across corpora (similar in one language but not the other). We selected the threshold combination that simultaneously minimized both quantities, yielding proportionally balanced and mutually consistent decisions across the two corpora. The search identified optimal thresholds of \SimThrCh~for Chinese and \SimThrEng~for English. \Cref{fig:SimilarityScatter} shows the distribution of similarity scores in both corpora and the pairwise relationship between Chinese and English similarity values, with the optimal thresholds marked for reference.

%% file: sections/appendix/ground_truth.tex
\section{Ground-truth annotation process}
\label{app:sec:annotation}

To ensure the reliability and transparency of our ground-truth dataset construction, we provide detailed documentation of the manual annotation process used to evaluate author name disambiguation performance.

\subsection{Annotation task design}
\label{app:sec:annotation:design}
Human annotators were presented with pairs of author names and their corresponding publication titles to determine whether the two publications were authored by the same individual. The primary annotation interface was designed as a structured spreadsheet with the following columns, though annotators had access to additional metadata when needed for disambiguation:

\begin{itemize}
    \item \textbf{name}: The author's name in Chinese characters as it appears in both publications
    \item \textbf{title1}: The title of the first publication (in Chinese)
    \item \textbf{title2}: The title of the second publication (in Chinese)  
    \item \textbf{manual\_check}: The annotation field where human judges record their decision (either ``match'' or ``not\_match'')
\end{itemize}

% \subsubsection{Sample Annotation Cases}

\Cref{app:tbl:annotation_examples} presents representative examples of the annotation cases shown to annotators. Each row represents one annotation task, where annotators must determine if the author name refers to the same individual across both publications and fill in the manual\_check column.

\begin{table}[H]
\centering
\caption{\textbf{Representative examples of the annotation cases presented to annotators.} Annotators were tasked with determining whether each pair of publications was authored by the same individual and recording their decision in the \textit{Manual Check} column. The column is shown empty as it appeared to annotators.}
\small
\begin{tabular}{p{1.5cm}p{5.5cm}p{5.5cm}p{1.8cm}}
\toprule
\textbf{Name} & \textbf{Title 1} & \textbf{Title 2} & \textbf{Manual Check} \\
\midrule
\zh{丁乾} & \zh{固体电解质与电极之间界面的分数维模型及其频率响应} & \zh{非晶态快离子导体电导特性的低频弛豫理论} &  \\
\midrule
\zh{丁菊仁} & \zh{镍基合金薄膜中的分形生长} & \zh{多层度分形理论及进展} &  \\
\midrule
\zh{万亚} & \zh{离子注入AlxGa1−xAs / GaAs和GaAs中的晶格损伤与相对化学} & \zh{1MeVSi+村底加温注入Al\_\_(0.3) Ga\_\_(0.7) As/GaAs超晶格和GaAs的晶格损伤研究} &  \\
\midrule
\zh{方贤绢} & \zh{新型钙钛矿铜氧材料Sr8 CaRe3 Cu4 O24的亚铁磁和轨道序性质} & \zh{MgNi2Bi4弹性和电子性质的第一性原理研究} &  \\
\midrule
\zh{万钧} & \zh{Cu表面弛豫和自扩散机制的修正嵌入原子法模拟} & \zh{掺铝SiOx的光致发光特性} &  \\
\midrule
\zh{何恰治} & \zh{纳米Ge颗粒镶嵌薄膜的Raman散射光谱研究} & \zh{六角结构金属中特殊位错组态的分析} &  \\
\bottomrule
\end{tabular}
\label{app:tbl:annotation_examples}
\end{table}

\subsection{Annotation guidelines}
\label{app:sec:annotation:guidelines}
Annotators were provided with the following guidelines for making disambiguation decisions:

\begin{enumerate}
    \item \textbf{Research Topic Consistency}: Consider whether both publications fall within plausible research interests of a single researcher.
    
    \item \textbf{Institutional and Temporal Coherence}: Examine publication dates and institutional affiliations to assess whether they represent a logical career progression (same institution, reasonable mobility patterns, or appropriate temporal sequence).
    
    \item \textbf{Conservative Matching Principle}: Annotators were instructed to use ``match'' only when there is clear evidence that two records belong to the same person. As long as there is any uncertainty or lack of clear evidence, cases should be marked as ``not\_match'' to minimize false positive matches.
\end{enumerate}

%% file: sections/appendix/confusion.tex
\section{Confusion matrices for disambiguation methods}
\label{app:sec:Confusion_Matrices}
To provide a comprehensive view of algorithm performance beyond aggregated metrics, we present detailed confusion matrices for all three disambiguation methods across both linguistic representations (\Cref{app:sec:Confusion_Matrices}).
 
Each confusion matrix shows the classification outcomes on our human annotated ground-truth dataset of \NsampleGT~author name pairs, where rows represent true labels and columns represent algorithmic predictions. The matrix (row-column) elements indicate: 
True Positive (match-match, correctly identified same individuals), 
False Negative (match-not match, missed matches where the algorithm failed to recognize the same person), 
False Positive (not match-match, incorrect merges where different individuals were erroneously classified as the same person), and 
True Negative (not match-not match, correctly identified different individuals). 

\Cref{app:sec:Confusion_Matrices}a displays results for the Chinese dataset, while \Cref{app:sec:Confusion_Matrices}b shows results for the English dataset. From left to right, the matrices correspond to Our Method, Sinatra et al.'s approach, and Waqas \& Qadir's framework. Comparing across methods, our approach demonstrates better recall performance while maintaining high precision, as evidenced by the higher True Positive counts and minimal False Positive errors. Notably, our method achieves zero False Positives on Chinese character names and only two on Pinyin names, indicating strong precision. The relatively lower False Negative counts (10 for Chinese, 9 for English) compared to alternative methods demonstrate improved sensitivity in detecting matching author pairs. Sinatra et al.'s method shows more conservative matching behavior with higher False Negative rates, while Waqas \& Qadir's approach exhibits the most conservative pattern with the highest False Negative counts across both datasets. 

\begin{figure}[H]
\includegraphics[width=1\textwidth]{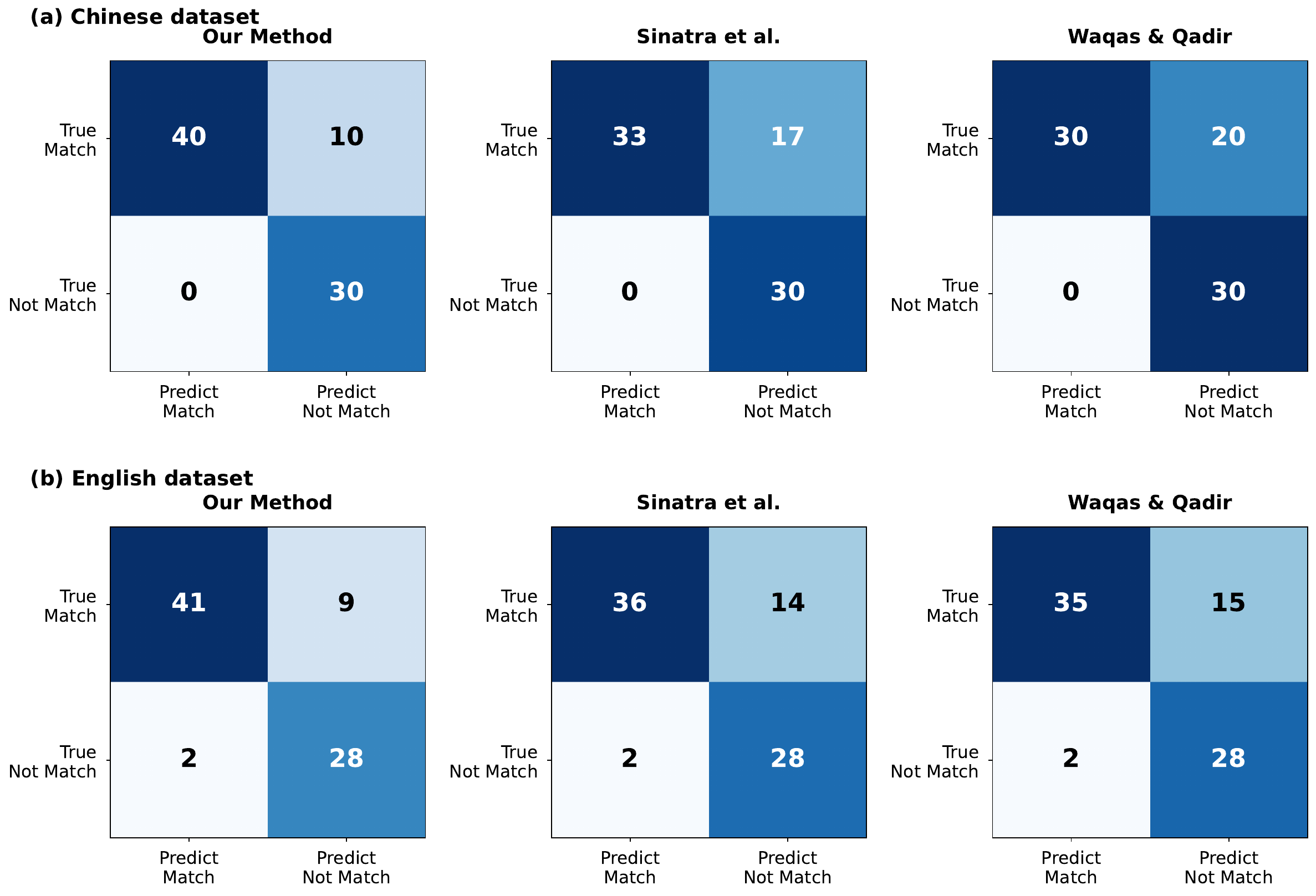}
\caption{\textbf{Confusion matrices of author name disambiguation across methods.} Classification outcomes for Our method, Sinatra et al.'s method, and Waqas \& Qadir's method on the Chinese dataset (a, top) and the English dataset (b, bottom). Each 2×2 matrix displays: True Positive (top-left), False Negative (top-right), False Positive (bottom-left), and True Negative (bottom-right) counts from manual validation of \NsampleGT~author name pairs. In this context, true positives refer to name pairs that represent the same individual and are correctly predicted as such, false positives are name pairs that actually refer to different individuals but are incorrectly predicted as the same, false negatives are true matches that the model fails to identify, and true negatives are correctly identified non-matching pairs.}
\label{app:fig:confusion}
\end{figure}